\documentclass[fleqn,usenatbib]{mnras}

\usepackage{newtxtext,newtxmath}
\usepackage[T1]{fontenc}
\usepackage{ae,aecompl}

\usepackage{graphicx}	% Including figure files
\usepackage{amsmath}	% Advanced maths commands
\usepackage{natbib} 
\usepackage{upgreek}
\usepackage{xcolor}
\usepackage{gensymb}
\newcommand{\cgs}{ erg\,cm$^{-2}$\,s$^{-1}$ }

\newcommand{\xmm}{{\it XMM-Newton}}
\newcommand{\xte}{{\it RXTE}}

\newcommand{\integral}{{\it INTEGRAL}}
\newcommand{\chandra}{{\it Chandra}}
\newcommand{\maxi}{{\it MAXI/GSC}}

\newcommand{\igr}{IGR J20155$+$3827}
\newcommand{\swift}{Swift J1713.4$-$4219}

\newcommand{\intg}{{\em INTEGRAL}}

\title[Multi-wavelength observations of X-ray binaries]{Multi-wavelength observations of the Galactic X-ray binaries IGR J20155+3827 and Swift J1713.4$-$4219}

\author[Onori et al.]{
F. Onori,$^{1,2}$\thanks{E-mail: francesca.onori@inaf.it}
M. Fiocchi,$^{1}$
N. Masetti,$^{3,4}$ 
A. F. Rojas,$^{5}$
A. Bazzano,$^{1}$ 
L. Bassani,$^{3}$
A.J. Bird$^{6}$
\\
$^{1}$Istituto di Astrofisica e Planetologia Spaziali (INAF), Via Fosso del Cavaliere 100, Roma, I-00133, Italy\\
$^{2}$INAF-Osservatorio Astronomico d'Abruzzo, via M. Maggini snc, I-64100 Teramo, Italy\\
$^{3}$INAF-Osservatorio di Astrofisica Spaziale e Scienza dello Spazio, via Gobetti 93/3, I-40129 Bologna, Italy\\
$^{4}$Departamento de Ciencias F\'isicas, Universidad Andr\'es Bello, Fern\'andez Concha 700, Las Condes, Santiago, Chile\\
$^{5}$Centro de Astronom\'ia (CITEVA), Universidad de Antofagasta, Avenida Angamos 601, Antofagasta, Chile\\
$^{6}$School of Physics and Astronomy, University of Southampton, University Road, Southampton, SO17 1BJ, UK
}

\date{Accepted XXX. Received YYY; in original form ZZZ}

\pubyear{2020}

\begin{document}
\label{firstpage}
\pagerange{\pageref{firstpage}--\pageref{lastpage}}
\maketitle

\begin{abstract}
In recent years, thanks to the continuous surveys performed by \integral\/ and {\it Swift} satellites, our knowledge of the hard X-ray/soft gamma-ray sky has greatly improved. As a result it is now populated with about 2000 sources, both Galactic and extra-galactic, mainly discovered by IBIS and BAT instruments. Many different follow-up campaigns have been successfully performed by using a multi-wavelength approach, shedding light on the nature of a number of these new hard X-ray sources. However, a fraction are still of a unidentified nature. This is mainly due to the lack of lower energy observations, which usually deliver a better constrained position for the sources, and the unavailability of the key observational properties, needed to obtain a proper physical  characterization. Here we report on the classification of two poorly studied Galactic X-ray transients \igr\/ and \swift\/, for which the combination of new and/or archival X-ray and Optical/NIR observations have allowed us to pinpoint their nature. In particular, thanks to \xmm\/ archival data together with new optical spectroscopic and archival Optical/NIR photometric observations, we have been able to classify \igr\/ as a distant HMXB. The new \integral\/ and {\it Swift} data collected during the 2019 X-ray outburst of \swift\/, in combination with the archival optical/NIR observations, suggest a LMXB classification for this source.  

\end{abstract}


%%%%%%%%%%%%%%%%%%%%%%%%%%%%%%%%%%%%%%%%%%%%%%%%%%

%%%%%%%%%%%%%%%%% BODY OF PAPER %%%%%%%%%%%%%%%%%%

\section{Introduction}
\label{sec:intro}

In the last fifteen years our knowledge of the hard-X/soft gamma-ray sky has greatly improved thanks to the continuous surveys performed with both \integral\/ \cite[][]{winkler03} and the Neil Gehrels Swift Observatory ({\it Swift}) \cite[][]{gehrels04} satellites, which have been in orbit since 2002 and 2004, respectively. To this aim, the main telescopes contributing to surveying are IBIS \cite[][]{ubertini03} on board of \integral\/ and BAT \cite[][]{barthelmy05} on board of {\it Swift}. These instruments are both operative in a similar energy range (about 15-200 keV) with a mCrab sensitivity limit and afford a point source location accuracy of a few arcmin (depending on the source strength, though IBIS provides a better angular resolution than BAT). As a result, the hard X-ray/soft gamma-ray sky is now populated with about 2000 sources, compared to the about 70 objects in the all-sky survey performed with the detectors of A4/HEAO1 \cite[in the range 13-180 keV at a flux sensitivity of 14 mCrab,][]{levine84} and the 40 sources reported from the sky images collected during 8 years of operation of the SIGMA imaging telescope on board the GRANAT satellite \cite[][]{revnivtsev04}. 

Indeed, the most recent IBIS survey \cite[][]{bird16} lists about 1000 sources, of which 307 are new detections, distributed mainly along the Galactic Plane, while the one reported by BAT \cite[][]{oh2018} contains about 1630 sources (422 new detections), 70\% of which are of extra-galactic origin, as this satellite was primarily designed to detect transient GRBs. 
A part of the newly reported detected sources have their nature still unidentified, about 20\% and $>$10\% for IBIS and BAT, respectively. This is mainly due to the transient nature of many of these sources, to the lack of a lower energy coverage which would allow a much better positional accuracy and to the lack of availability of key observational properties (such as spectral shape, flux, variability and absorption) which are needed to properly characterize these sources.

Many different follow-up campaigns have been performed by several groups, especially for the new sources discovered by IBIS (IGRs) and BAT \cite[see][and references therein]{tomsick08, rodriguez10,bernardini13, tomsick16, landi17, tomsick20}. These campaigns make use of archival data or new data-sets in the X-ray band with \xmm, \chandra\/ and XRT/{\it Swift} instruments, in order to detect the soft X-ray emission from these sources and thus drastically reduce the positional uncertainty down to (sub-)arcsecond size. Subsequently, optical and/or near-infrared (NIR) spectroscopy is performed on their putative lower-energy counterpart(s) with the aim to uncover their nature \cite[see e.g.][and references therein, to name a few]{rahoui08, masetti13, parisi14, rojas17, fortin18, karasev18, marchesini19}. As a result, the nature of more than 300 \integral\/-discovered sources (IGRs) have been identified on a total of 560 listed in \citet{bird16} -- mainly active galactic nuclei, cataclysmic variables (CV), but also high-mass X--ray binaries (HMXB) and low-mass X--ray binaries (LMXB) and a similar number is available for the new BAT transients, though the latter are dominated by extra-galactic sources.   

In this paper we report on the possible classification for two poorly studied transient sources: \igr\/ and \swift. For each of these objects we first briefly report on data presented so far in the literature, then we focus on new data obtained with \integral, \xmm\/ and {\it Swift} and finally we use archival and newly-acquired optical/NIR observations, in order to attempt to pinpoint their nature. In Table \ref{tab:obs_log} the details of the observations available for both sources are shown.

All the {\it Swift} data reported throughout this manuscript have been processed by using the standard tools incorporated in \texttt{HEAsoft} (v6.26.1), while the the spectral analysis is always performed by using the software \texttt{XSPEC} v12.10.1 \cite[]{Arnaud96}.

\begin{table}
\caption{List of the \igr\/ and \swift\/ observations. (1) Date of observation; (2) instrument; (3) observational setup (4) exposure time}
\label{tab:obs_log}
\begin{tabular}{llll}
\hline
MJD			& Instrument & Obs. setup & Exp. time  \\
(days)      &            &                     & (s)             \\
(1)         & (2)        & (3)                 & (4)             \\
\hline
\multicolumn{4}{c}{\igr}\\
\hline
57874.39    & EPIC/\xmm  & Medium filter &  22400  \\
58746.90    & BFOSC/Cassini & Grism \#4; slit 2$''$  &  2$\times$1800 \\
\hline
\multicolumn{4}{c}{\swift}\\
\hline
58148.89    & XRT/{\it Swift} & Photon Counting & 1421.64\\
            & UVOT/{\it Swift} & $U$   & \phantom{1}836.62\\
58149.82    & XRT/{\it Swift} & Photon Counting & 1923.09\\
            & UVOT/{\it Swift} & $UVW2$ &\phantom{1}930.07\\
58759.70    & IBIS/\integral   & Photon-by-Photon & 4200.00$^{*}$ \\
            & JEM-X/\integral  & Photon-by-Photon & 7800.00$^{*}$ \\
58779.04    & XRT/{\it Swift} & Photon Counting & 1465.90\\
            & UVOT/{\it Swift} & $UVW1$ & 1456.74\\
58785.81    & XRT/{\it Swift} & Windowed Timing & 2743.48\\
            & UVOT/{\it Swift} & $UVW2$  & 1669.95\\
\hline
\end{tabular}
$^{*}$Effective exposure times
\end{table}

\section{IGR J20155+3827}
\label{sec:igr}
\begin{figure}
\includegraphics[angle=0, width=1\columnwidth]{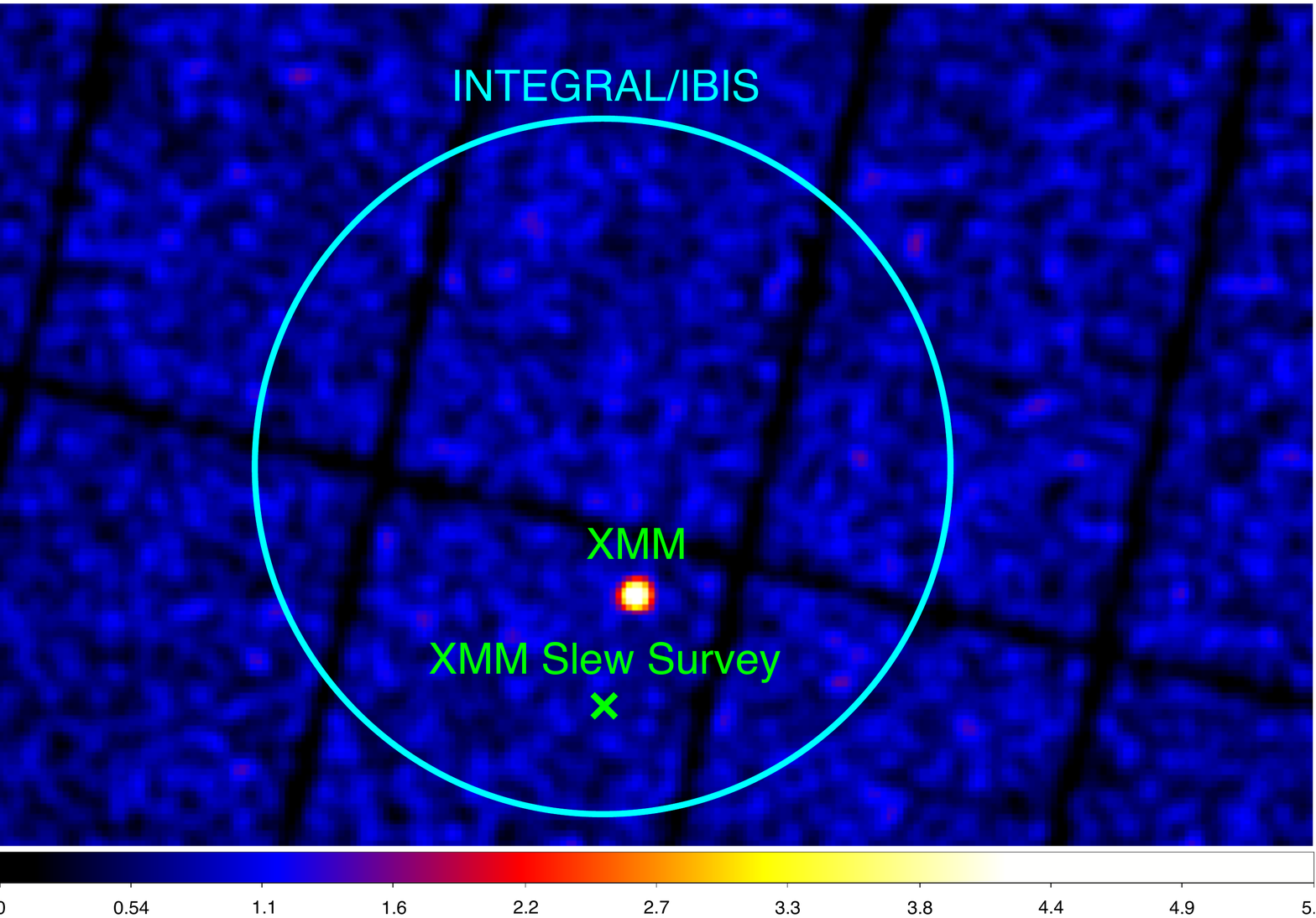}
\caption{ Image of the \igr\/ field in the 4.5-12 keV band from the {\it XMM}/EPIC observation taken on MJD 57874.39. The green cross indicates the position of the X-ray transient detected in the XMM slew survey on MJD 55686. The cyan circle shows the {\it INTEGRAL}/IBIS position of \igr\/ with a a 90\% confidence radius of 4.2 arc minutes}
\label{fig:igrxmmfield}
\end{figure}

\igr\/ was first reported in the 4th IBIS/{\it INTEGRAL} catalog \cite[][]{bird2010} as new transient X-ray source. It was detected by using the "bursticity" method \cite[see][]{bird2010} at a peak flux of 5.8 mCrab in the 17-30 keV band, during a 25.5 day outburst, starting on 2004 September 08 (MJD=53256.10). 
The position of the source as derived from the IBIS image analysis is RA (J2000) = 20:15:30 and DEC (J2000)= +38:27:00, within a 90\% confidence radius of 4.2 arc minutes. The source was later confirmed in the 1000 orbit catalog, by using a much longer exposure time of 3.4 Ms versus 1.3 Ms \cite[][]{bird16}. \igr\/ was also detected by \xmm\ on 2017 May 01 (MJD 57874.39), which was able to deliver a better  constrained position of RA (J2000) = 20:15:28.03 and DEC(J2000) = +38:25:28.6, with a 90\% confidence error radius of 1$\farcs$1. The detection of this source is also reported in the 4XMM$-$DR10 catalog \cite[][]{webb20} with an observed flux in the 0.2-12 keV band of F$_{0.2-12}$=(2.4$\pm$0.2)$\times10^{-13}$ \cgs. 
We note that the same field has been previously observed on 2011 May 05 in the framework of the \xmm\/ slew survey \cite[][]{saxton08}. The authors reported the detection of a source at the position RA (J2000) = 20:15:29.9 and DEC (J2000) = 38:24:06.5, which is inside the \integral\,/IBIS 90\% confidence radius for \igr\/, but it is not compatible with the \xmm\ position derived during the 2017 May 01 observation (see Figure \ref{fig:igrxmmfield})\footnote{We note however that the XMSLew position may be effected (flagged) by attitude reconstruction problems which provide an uncertainty on the source position of up to 1 arcmin, making the likelyhood of an association still possible.  If the two XMM sources coincide, then the change in flux over a six years period is of a factor of 15.}. Moreover, the source reported by the \xmm\/ slew survey is characterized by a soft emission in the 0.2-2 keV channel of F$_{0.2-2}$=(1.55$\pm$0.46)$\times$10$^{-12}$\cgs, while the flux measured in the total band is F$_{0.2-12}$=(3.67$\pm$1.10)$\times$10$^{-12}$\cgs. No flux measurements are reported for the 2-12 keV energy channel. Instead, the \xmm\/ and \integral\/ observations of \igr\/ indicate that this source is characterized by an hard emission. In particular, the \igr\/ spectrum extracted from the \xmm\/ observations of MJD 57874.3 show that the source emission is mainly detected in the 2-10 keV band (see Figure \ref{fig:igrspec}). Thus, we tentatively  exclude that the transient reported by the \xmm\/ slew survey is related to \igr. 

\begin{figure}
   \includegraphics[angle=0, width=1\columnwidth]{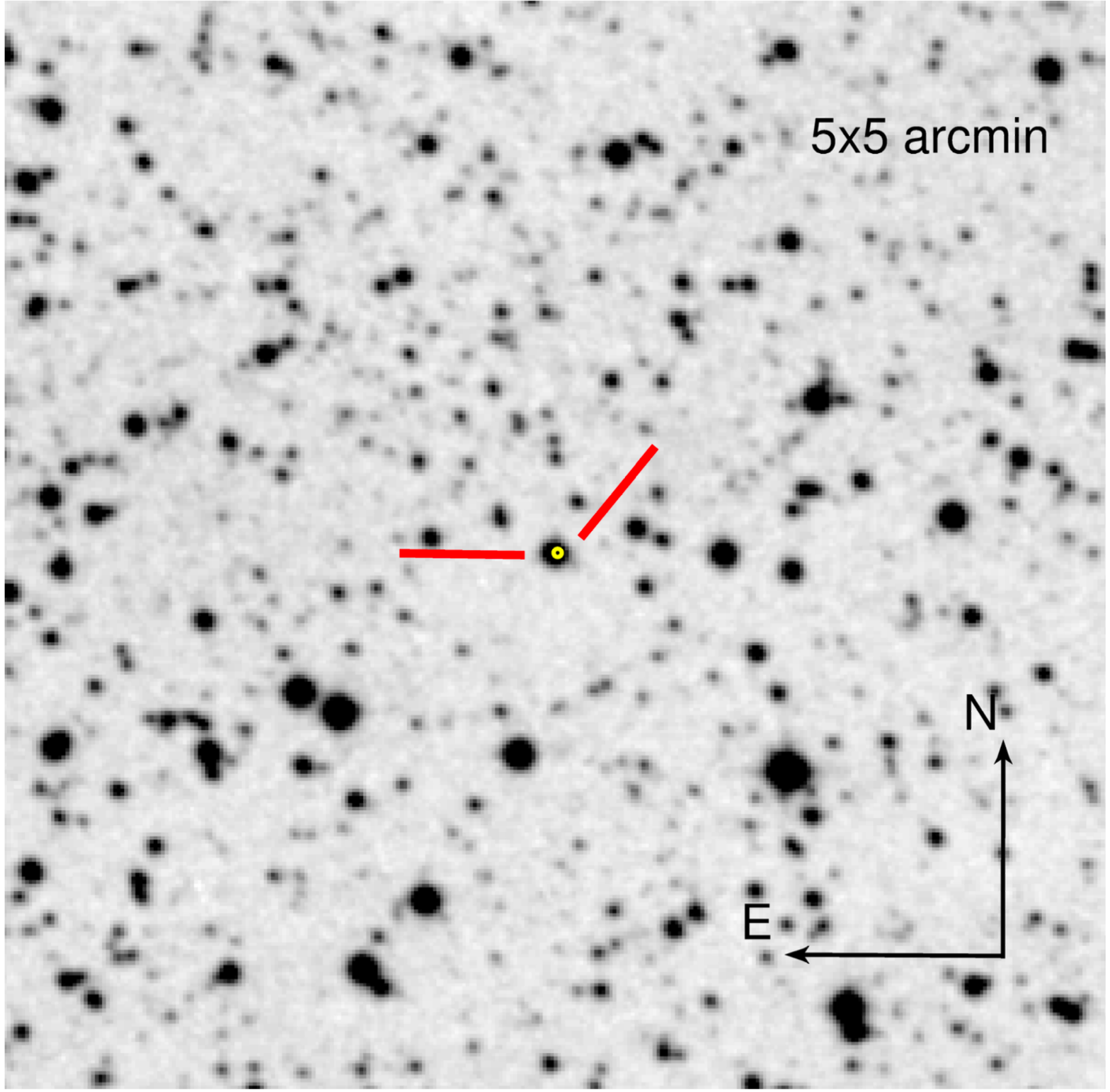}
   \caption{Optical field of \igr\/ extracted from the DSS-II-Red survey. Size and orientation are reported in the figure. The \xmm\/ 90\% confidence level error region of 1$\farcs$1 radius is shown as a yellow circle at the center of the image. One single optical source, USNO$-$B1 1284$-$0410895 (indicated by the red tick marks)}, lays inside it.
\label{fig:igr_fc}
   \end{figure}

We have identified an optical/infrared counterpart candidate for \igr\/ with the star USNO$-$B1 1284$-$0410895\footnote{Other names of the source are 2MASS J20152803+3825260 and ALLWISE J201528.03+382526.0} \cite[][]{monet03}, at the position of RA(J2000) = 20:15:28.041, DEC(J2000) = +38:25:25.27. This is the only object laying within \xmm\/ position error radius for \igr\/, as shown in figure \ref{fig:igr_fc}. 
%USNO$-$B1 1284$-$0410895 is also reported in the {\it Gaia} Data Release 2 (DR2) catalog \cite[][]{gaia16, gaia18, riello18} and a 
%distance of d=(10$\pm$6) kpc can be derived through the parallax measurements reported in the {\it Gaia} database.
USNO$-$B1 1284$-$0410895 is reported in the {\it Gaia} Data Release 2 (DR2) catalog \cite[][]{gaia16, gaia18, riello18}, where a 
distance d = 10$\pm$6 kpc can be derived through the small parallax measurement value of 0.0996$\pm$0.0589 mas. The source is also reported in the more recent {\it Gaia} Early Data Release 3 (EDR3) catalog \cite[][]{gaia20,riello20,fabricius20} where an even smaller parallax measurement of 0.0795$\pm$0.0177 mas is delivered. 

However there are some issues regarding the distance estimate from the {\it Gaia} parallax. First, according to \cite{bailer2020} and \cite{bailer2018}, reliable distances cannot be obtained by inverting the parallax for the majority of stars in {\it Gaia} catalogs. Secondly, the parallax value delivered for USNO$-$B1 1284$-$0410895 in {\it Gaia} DR2 is affected by an astrometric excess noise at a level of 0.474 mas, indicating that the fit for the parallax determination is not necessarily reliable. We note that the astrometric excess noise is considerably reduced to a level of 0.149 mas in {\it Gaia} EDR3, indicating a more reliable parallax fit with respect the DR2 estimate. Interestingly, distances for a number of {\it Gaia} DR2 and EDR3 stars have been inferred by applying the probabilistic analysis described in  \cite{bailer2018} and \cite{bailer2020}, respectively. For USNO$-$B1 1284$-$0410895 a geometric distance d = 5.6$\pm$1.5 kpc has been derived by using {\it Gaia} DR2 data, while a geometric distance d = 7.2$\pm$1.0 kpc has been obtained from the more accurate EDR3 data. Thus, we assume a distance of d$\sim$7 kpc as the most reliable for the optical counterpart of \igr.

The identification of an optical counterpart candidate and the derived distance from parallax measurements indicate that \igr\/ is a Galactic transient.
In the following we report on new results obtained with \xmm\/ and optical data analysis.

\subsection{{\it XMM-Newton} observation of \igr}
\label{sec:XMMigr}

\begin{figure}
\includegraphics[width=1\columnwidth]{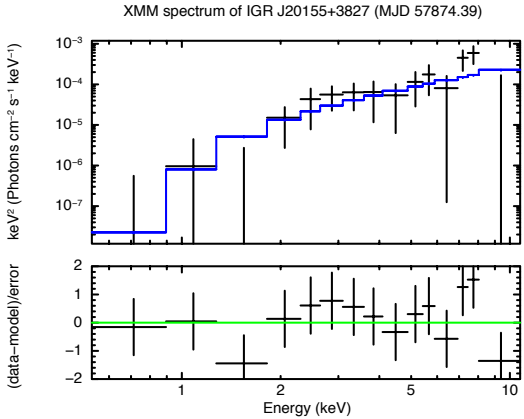}
\caption{ {\it XMM}/EPIC PN unfolded spectrum of IGR J20155$+$3827 extracted from the MJD 57874.39 observation (upper panel), together with the residuals with respect to the best-fit model (lower panel). The absorbed power-law model is shown as blue solid line over-plotted on the data points in the upper panel.}
\label{fig:igrspec}
\end{figure}

We processed the \igr\/ \xmm\/ observation performed on 2017 May 01 (MJD 57874.39, see Table \ref{tab:obs_log} for more details) using the {\it XMM} Science Analysis System \texttt{SAS v16.1.0}. 
Unluckily, the pointing was affected by a background flare and,  consequently, to avoid its impact on our analysis, we applied a filter in time considering only the first  1.4$\times$10$^4$ s of the observation when extracting the PN and MOS spectra. 
Since the MOS data result in a poor signal-to-noise ratio, we report only the spectrum extracted from the PN data, which cover a broader energy range (0.2-10 keV) with a better sensitivity. The data can be modelled by an absorbed power-law (\texttt{PHABS*POWERLAW}), as shown by the residuals with respect to the model reported in Figure \ref{fig:igrspec}, bottom panel. 
As a result, we obtain a high value for the hydrogen column density with a wide error range: N$_{\mathrm{H}}$=(5.0$^{+10.4}_{-4.2}$)$\times$10$^{22}$ cm$^{-2}$. We also obtain a spectral index $\Gamma$= 1.8$^{+3.4}_{-1.7}$ and $\chi^{2}/$d.o.f. of 7.92/12. We derived an upper limit for the unabsorbed flux in the 0.2-12 keV band, which is F$_{0.2-12} \leq$1.2$\times$10$^{-10}$\cgs.
Given the poorly constrained value of the N$_{\mathrm{H}}$ parameter obtained in this way, we performed the fitting procedure with the hydrogen column density parameter fixed to the Galactic value N$_{\mathrm{H}}$=1.11$\times$10$^{22}$ cm$^2$, as derived for the source position by using the Heasarch \texttt{nH} \texttt{ftools} and the HI4PI Collaboration map \cite[][]{HI4PI}. 
In this case, we obtain a spectral index of $\Gamma$=0.4$^{+0.7}_{-0.9}$, a $\chi^{2}/$d.o.f. of 10.11/13 and an unabsorbed flux in the 0.2-12 keV band of F$_{0.2-12}$=(3.55$\pm$1.80)$\times$10$^{-13}$\cgs.

In Table \ref{tab:igr_pnspec} the results obtained from both the spectral fitting procedures are reported.
We remark that the use of more complex models
\cite[e.g., a power law with a high-energy cut-off;][]{white83} is not justified statistically and returns largely undetermined fit parameters. Finally, no periodicities have been found from the timing analysis performed on the \xmm\/ data in the 0.2-12 keV energy range.

\begin{table}
\caption{Results from the spectral analysis on \igr\/ \xmm\/ observation of MJD 57874.39.}
\label{tab:igr_pnspec}
\centering
\begin{tabular}{lll}
\hline\hline
Model				        &Parameter		   & Value \\
\hline			
\hline
\texttt{PHABS}              &N$_{\mathrm{H}}$($\times$10$^{22}$ cm$^{-2}$)  & 5.0$^{+10.4}_{-4.2}$ \\
\texttt{POWERLAW}           &$\Gamma$ 		   &1.8$^{+3.4}_{-1.7}$ \\
                            & norm             &  $\leq$1.7$\times$10$^{-2}$ \\
                            &$\chi^{2}$/d.o.f.                  &7.92/12\\
                           \hline
                           \hline
\multicolumn{3}{c}{$^{*}$Flux in the 0.2-12 keV band: F$_{0.2-12}\leq$1.2$\times$10$^{-10}$ \cgs}\\
\hline
\texttt{PHABS}              &N$_{\mathrm{H}}$($\times$10$^{22}$ cm$^{-2}$)     & 1.11 (frozen) \\
\texttt{POWERLAW}           &$\Gamma$ 		   &0.4$^{+0.7}_{-0.9}$ \\
                            & norm             & $\leq$1.8$\times$10$^{-5}$ \\
                            &$\chi^{2}$/d.o.f.                  &10.11/13\\
\hline
\multicolumn{3}{c}{$^{*}$Flux in the 0.2-12 keV band: F$_{0.2-12}$=(3.55$\pm$1.80)$\times$10$^{-13}$ \cgs}\\
\multicolumn{3}{l}{calculated by freezing the normalization parameter.}\\
\hline
\hline
\end{tabular}
$^{*}$Fluxes are unabsorbed\\
\end{table}

\begin{figure}
\includegraphics[width=\columnwidth]{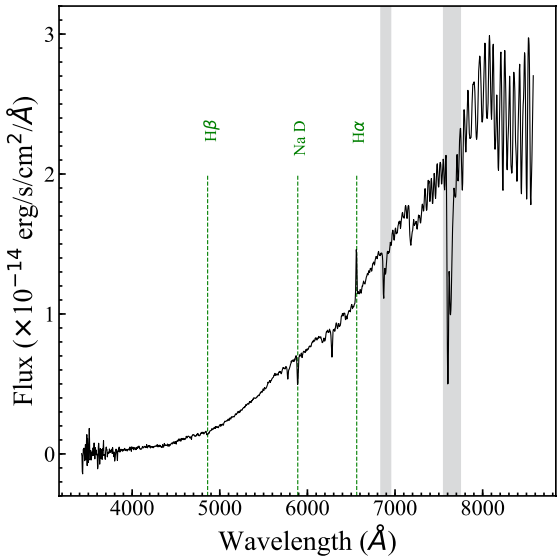}
\caption{Optical spectrum (not corrected for the intervening Galactic absorption) of the counterpart of IGR J20155+3827. The main spectral features are labelled. Grey areas indicate the position of telluric absorption bands due to atmospheric O$_2$.}
\label{fig:fig3}
\end{figure}

\subsection{Optical spectroscopy}
\label{sec:OPTigr}
 
We spectroscopically observed the putative optical counterpart of \igr, USNO-B1 1284-0410895, with the BFOSC instrument mounted on the 1.5m `G.D. Cassini' telescope of the INAF-OAS in Loiano (Italy), equipped with a 1300$\times$1340 pixels EEV CCD. The main information on these data are shown in Table \ref{tab:obs_log}. In particular, the observation were carried out on 2019 September 20, starting at 21:31 UT (MJD 58746.90). Two 1800s exposures were acquired with the grism \#4 and using a slit width of 2$''$. This observational setup secured a dispersion of 4.0 \AA/pixel. 

The observations were bias corrected, flat-fielded, cleaned for
cosmic rays, background-subtracted and the corresponding spectra
were extracted following standard procedures \cite[][]{horne86} using
IRAF\footnote{IRAF is the Image Reduction and Analysis Facility, distributed by the National Optical Astronomy Observatories, which are operated by the Association of Universities for Research in Astronomy, Inc., under cooperative agreement with the National Science Foundation. It is available at {\tt http://iraf.noao.edu}.}.
Wavelength calibration was performed with the use of Helium-Argon lamp acquisitions,  taken after each science data frame. The final uncertainty in the wavelength calibration was of 3 \AA. The object spectra were then flux-calibrated by means of the IRAF spectro-photometric standard star BD +26 2606 which was observed with the same setup described above. The two spectra of USNO-B1 1284-0410895 were eventually stacked together to increase the final signal-to-noise ratio (see Fig. \ref{fig:fig3}).

The analysis of the spectrum of the putative optical counterpart of IGR J20155+3827 indicates the presence of a narrow H$_\alpha$ emission line consistent with its rest-frame wavelength superimposed on a very reddened continuum, likely affected by Galactic absorption along the line of sight of the source (see Section \ref{sec:XMMigr}). Other apparent features in absorption correspond to the rest-frame H$_\upbeta$, the interstellar \ion{Na}{} doublet at 5890 \AA\/ and diffuse interstellar bands at $\lambda\lambda$ 5780, 6280 \AA. We measure flux and equivalent width (EW) of the H$_\upalpha$ emission as F$_{\rm H\upalpha}$=(5.0$\pm$0.4)$\times$10$^{-14}$ erg cm$^{-2}$ s$^{-1}$ and EW = (4.5$\pm$0.4) \AA, respectively.

\section{Swift J1713.4$-$4219}
\label{sec:swift}

The X-ray transient \swift\ was first detected by {\it Swift}/BAT on 2009 Nov 13 (MJD 55148) in the 15-50 keV band at position RA (J2000) = 17:13:26.6 and DEC (J2000) = -42:19:37.2, with a 90\% confidence error radius of 3.0 arc minutes \cite[][]{krimm09}. On 2009 Nov 16 (MJD 55151), the transient was observed with \xte\/ and the  spectral properties of the \xte\,/PCA spectrum in the 2.5-30 keV band were interpreted as consistent with a black hole transient in the low-hard state \cite[][]{krimm09}.  
A renewed phase of activity of \swift\ was observed during the \intg\/ Galactic Plane  Program scans of the Norma region on 2019 Oct 3, revolution 2144 \cite[MJD 58759.70, ][]{onori19}. The observations were executed between 2019-10-03 16:52:57 UTC and 2019-10-04 09:20:32 UTC. 
The source was detected by the IBIS/ISGRI instrument both in the 22-60 keV and 60-100 keV energy bands and by the JEM-X instrument, but only below 10 keV. 
Subsequent XRT/{\it Swift} observations have been executed on 2019 October 23 (MJD 58779.04), i.e. about 20 days after the \integral\/ detection of the source, reporting the detection of the transient in the 0.3-10 keV band \cite[][]{baglio19}. 

An optical counterpart was pinpointed among three possible candidates thanks to the Las Cumbres Observatory (LCO) $i^{\prime}$-band images of the field \cite[][]{baglio19} together with the \chandra\ refined localization of the transient \cite[][]{chakrabarty19}, which led to the identification of the \swift\ optical counterpart with the star A of the LCO observations  \cite[see][for more details]{baglio19}. The reported position of the star is RA (J2000) = 17:13:40.993 and DEC (J2000)= -42:18:38.20. No counterpart has been detected in the {\it Swift}/UVOT observation in the $UVW1$ band.
Finally, only an upper limit of 115 $\mu$Jy was reported by radio measurements performed with MeerKAT on 2019 Oct 26 \cite[][]{girard19}.

In the following, we report on the results from the observations performed on the new \integral\/ and {\it Swift} data collected during the 2019 X-ray outburst of \swift\/ and from the optical/NIR archival information available for the optical counterpart candidate.

\subsection{The X-ray outburst of \swift}

\begin{figure}
   \includegraphics[width=1\columnwidth]{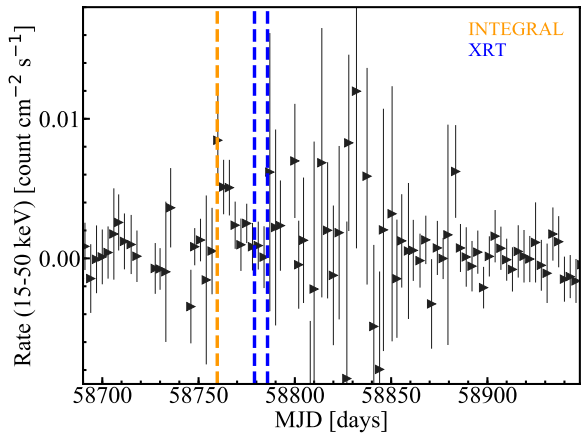}
   \caption{{\it Swift}/BAT light curve of \swift\/ with a binning of 3 days. The epochs of the \integral\/ and XRT/\textit{Swift} observations are indicated by yellow and blue vertical dashed lines, respectively.}
   \label{fig:BAT_lc}
\end{figure}

The {\it Swift}/BAT light curve of \swift\/ starting from MJD 58700 is shown in Figure \ref{fig:BAT_lc}, where the epochs of \integral\/ and  XRT observations are marked with yellow and blue vertical dashed lines, respectively. While the \integral\/ observations of the source (MJD 58759.70) coincide with the peak emission of the hard X-ray outburst, the {\it Swift}/XRT pointings have been executed at later epochs, when it was fading. We note that the source is not monitored in the \maxi\ survey so that the only soft X-ray band  observation  simultaneous with the IBIS data is derived from the JEM-X instrument. 
 
\begin{figure*}
   \includegraphics[width=1.0\columnwidth]{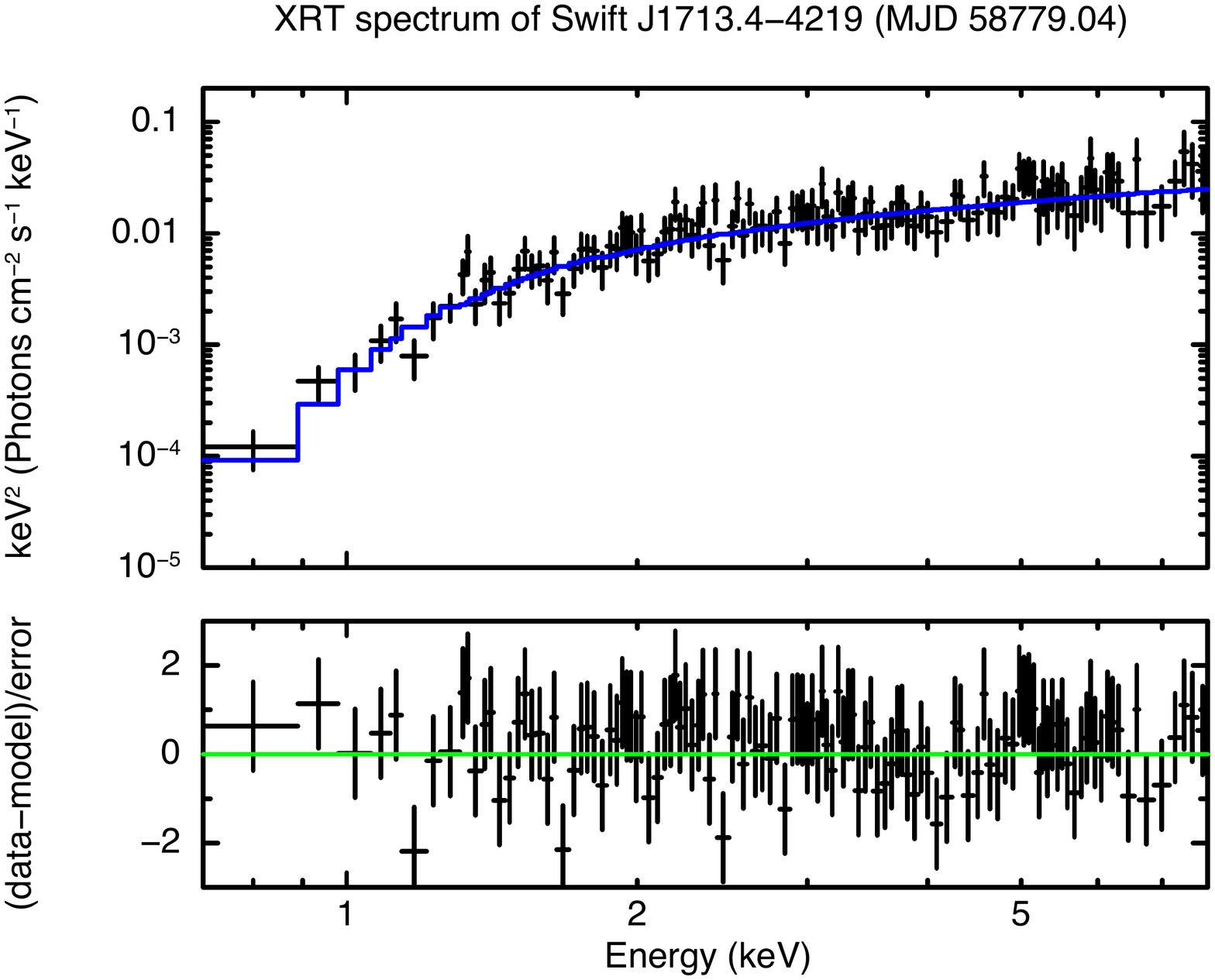}
   \includegraphics[width=1.0\columnwidth]{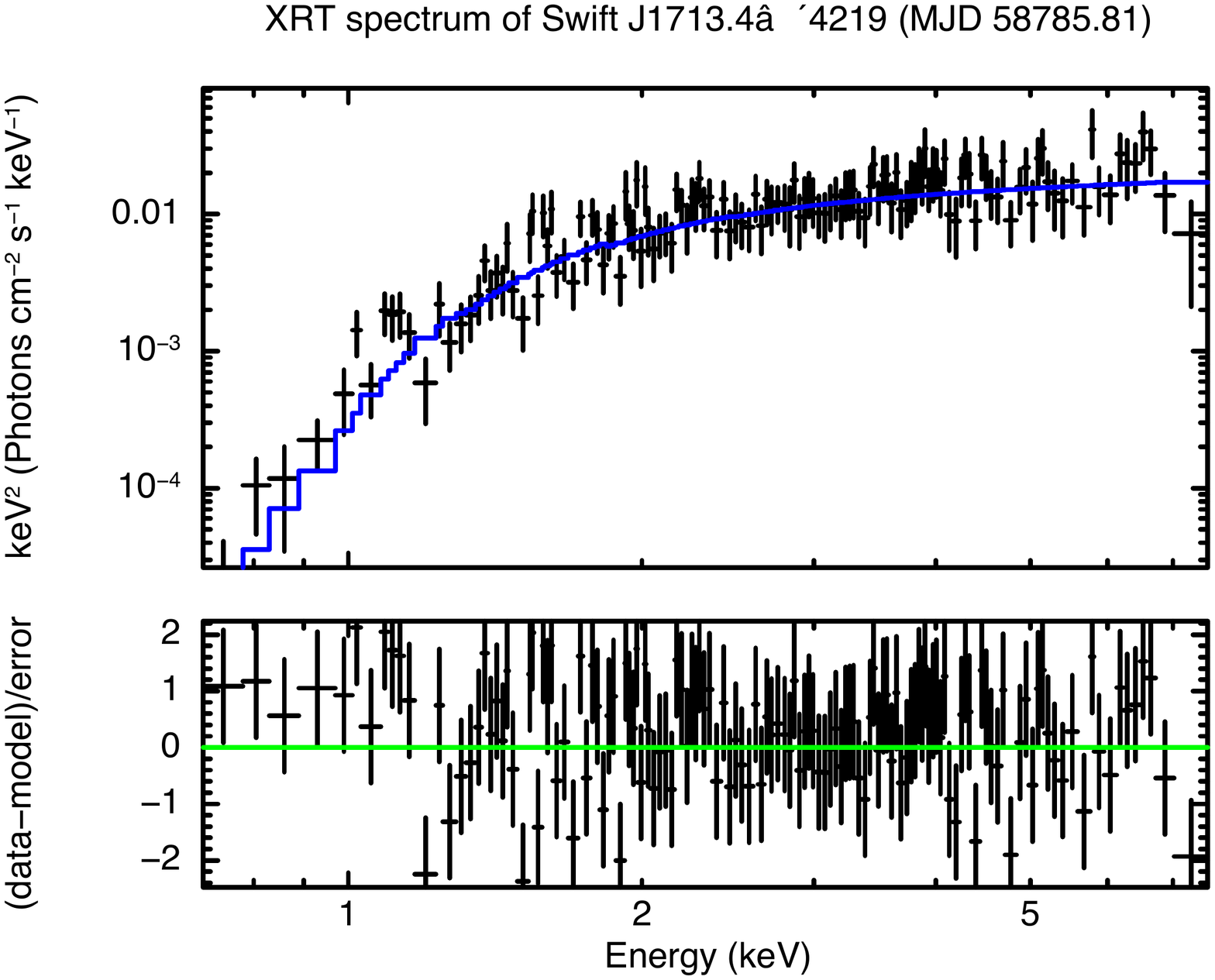}
   \caption{{\it Swift}/XRT unfolded spectra of \swift\/  extracted from observations taken on MJD 58779.04 (left panel) and on MJD 58785.81 (right panel). Both spectra have been modelled with an absorbed powerlaw. The residuals with respect the model are shown in the bottom panels. The spectrum obtained from data taken on MJD 58785.81 shows a soft excess at energies $\leq$10 kev and the presence of the iron line at 6.5 keV. }
   \label{fig:xrt_spec}
\end{figure*}

\subsubsection{The {\it Swift}/XRT data}
\label{sec:xrtspec}

Starting from its first discovery, \swift\/ has been observed by the {\it Swift}/XRT instrument in different epochs, for a total of four observations, as shown in Table \ref{tab:obs_log}. The first three pointings were taken in photon counting mode, while the last one was executed in windowed timing mode.

We do not detect emission from \swift\/ in the first two XRT observations (MJD 58148.89 and MJD 58149.82), while it is clearly detected in the data  on MJD 58779.04 and MJD 58785.81, about 20 days after the \integral\/ detection of the renewed activity of source in 2019. From these data we extracted the source spectrum by using a circular region of radius $\sim$50\arcsec\, centered on the source position and a background circular region of radius $\sim$140\arcsec\, placed in an area free of sources. In Figure \ref{fig:xrt_spec} the spectra obtained from the observations taken on  MJD 58779.04 (left panel) and MJD 58785.81 (right panel) are shown. In particular, the spectrum of MJD 58779.04 is well represented by the model \texttt{PHABS*POWERLAW}, as shown by the residuals (Fig. \ref{fig:xrt_spec}, left side, bottom panel). In this case, we obtain  N$_{\mathrm{H}}$=(1.0$\pm$0.3)$\times$10$^{22}$  cm$^{-2}$, the spectral index $\Gamma$=1.4$\pm$0.3 and a $\chi^{2}$/d.o.f = 75.85/93. The unabsorbed flux in the 0.2-10 keV energy band is F$_{\mathrm{0.2-10}}$=7.8$\times$10$^{-11}$ \cgs. In Table \ref{tab:Swift1713fit} we report the best fitting parameters obtained from this spectral analysis. These values are fully compatible with what has been  found by \cite{baglio19}, who interpreted these X-ray properties within an  X-ray binary system in the hard state scenario. 

When applying the same model to the subsequent spectrum of MJD 58785.81, a soft excess at energies $\leq$1.0 keV together with the iron line around 6.4 keV emerge from the residuals (bottom panel in Figure \ref{fig:xrt_spec}, right side). These spectral properties are indicative of transition of the system from a high state in correspondence of the \integral\/ observation towards a soft state. This is also supported by the location of the XRT observations in the BAT lightcurve, shown in Figure \ref{fig:BAT_lc} with blue dashed lines, with respect to the location of the \integral\/ observation (yellow dashed line in Figure \ref{fig:BAT_lc}). While during the first XRT observation (corresponding to MJD 58148.89), the hard X-ray outburst was still declining, the XRT observation on MJD 58149.82, where the soft excess and the iron line around 6.4 keV appear, falls just at the end of the hard X-ray outburst. 

\begin{figure}
   \includegraphics[width=1\columnwidth]{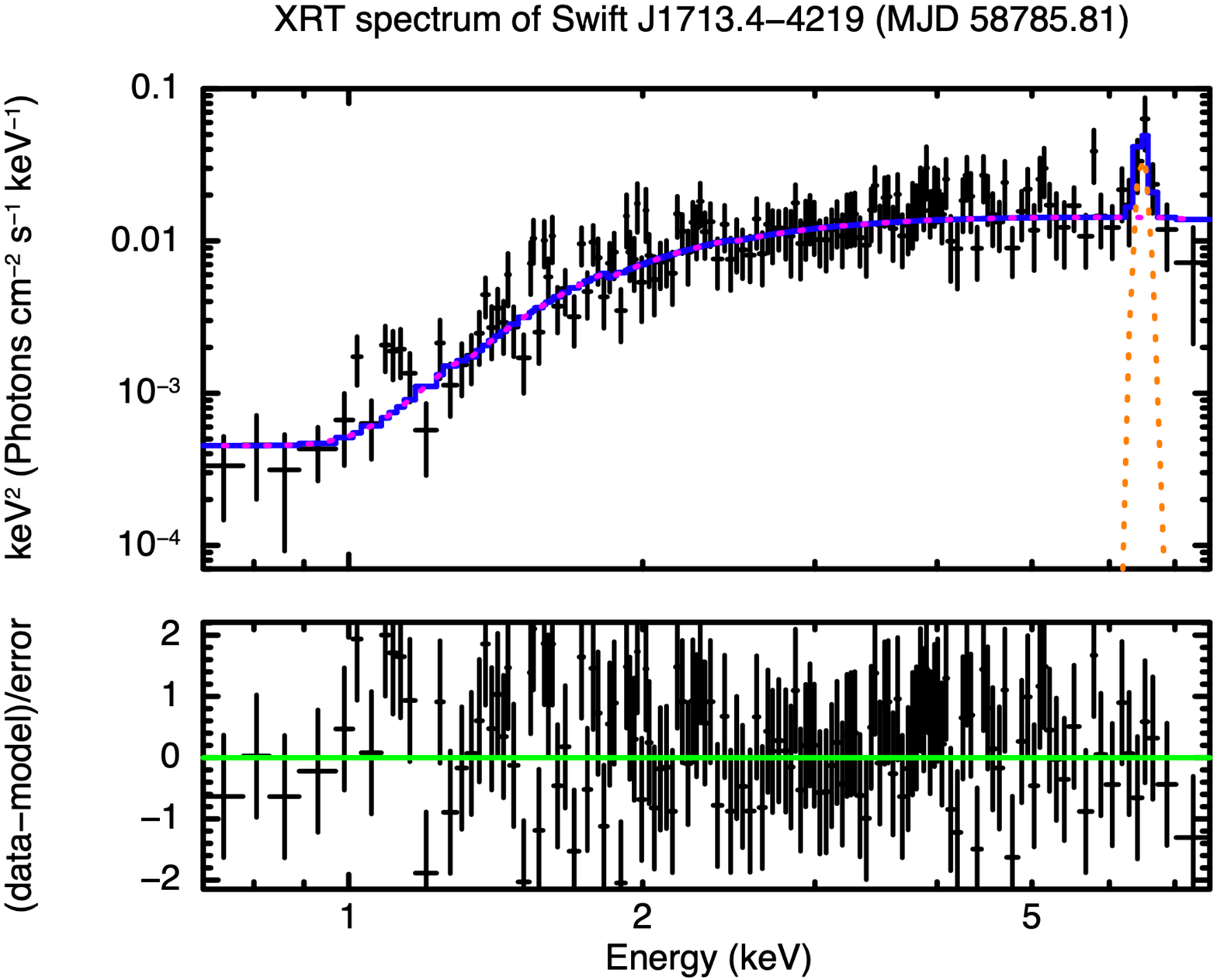}
   \caption{Spectral analysis of the \swift\/ spectrum taken on MJD 58785.81 using the model \texttt{PCFABS*(POWERLAW+GAUSS)}. The residuals with respect the model are shown in the bottom panel. Solid blue line indicate the full model, while colored dashed lines show each component (\texttt{GAUSS} with orange and \texttt{POWERLAW} with magenta.)
   }
   \label{fig:xrt_spec2}
\end{figure}

In order to model the soft excess and the iron line observed on MJD 58785.81 spectrum, we have analyzed it using the model \texttt{PCFABS*(POWERLAW+GAUSS)}. In Figure \ref{fig:xrt_spec2} we show the spectral analysis of the MJD 58785.81 spectrum (upper panel) together with the residuals with respect to this model, shown by the solid blue line in the upper panel. The single components are shown with dashed lines of different colors.
In Table \ref{tab:Swift1713fit} the best fitting parameters from this spectral analysis are reported.
In particular, we obtain N$_{\mathrm{H}}$=(2.1$\pm$0.2)$\times$10$^{22}$  cm$^{-2}$, a covering fraction of 0.98$\pm$0.01, a photon index $\Gamma$=2.1$\pm$0.1 and a $\chi^{2}$/d.o.f = 129.25/132. The iron line is narrow and centered at 6.5$\pm$0.2 keV with an equivalent width (EW) of  EW=0.58$\pm0.35$ keV. Unfortunately, only an upper limit is derived for the line width ($\sigma_{\mathrm{Fe}}\leq$0.30 keV). 
The unabsorbed flux in the 0.2-10 keV energy range is F$_{0.2-10}$= 1.2$\times$10$^{-10}$\cgs.

\begin{table*}
\caption{Results from the spectral analysis of Swift J1713.4-4219. Different columns correspond to: (1) Component used in the model; (2) component parameters; (3) Values of the parameters for the spectral analysis of JEM-X+IBIS data; (4) and (5) Values of the parameters for the spectral analysis of XRT data. We report the value for the unabsorbed flux in the 0.2-10 keV and 3-200 keV bands, for XRT and JEM-X+IBIS data, respectively.}
\label{tab:Swift1713fit}
\centering
\begin{tabular}{lllll}
\hline\hline
Model				        &Parameter			&JEM-X+IBIS      & XRT		     & XRT	         \\
         			        &                	&MJD 58759.70    & MJD 58779.04	 & MJD 58785.81	  \\
(1)         			    & (2)               & (3)           & (4)               & (5)          \\
\hline
\multicolumn{5}{c}{\texttt{PHABS*{POWERLAW}}}\\
\hline
&&&&\\
\texttt{CONSTANT}           & $C$(JEM-X)       & 0.78$^{+0.80}_{-0.40}$ & $\cdots$ & $\cdots$ \\
                            & $C$(IBIS)        & 1.0 (frozen)   & $\cdots$ & $\cdots$\\
\texttt{PHABS}              &N$_{\mathrm{H}}$($\times$10$^{22})$ cm$^{-2}$ &1.04 (frozen) &1.04$\pm$0.30	&1.38$\pm$0.35\\
\texttt{POWERLAW}           & $\Gamma$               	                  &1.90$\pm$0.30  &1.40$\pm$0.30	 & 1.75$\pm$0.27  \\     
                            & norm                   	                  &$\leq$0.44      &0.008$\pm$0.004	& 0.011$\pm$0.005 \\
                            &$\chi^{2}/$d.o.f.                            &3.27/7  	&75.85/93		    & 175.12/148\\
&\\
\multicolumn{2}{l}{Unabs. Flux ($\times$10$^{-10}$) \cgs}  & $\leq$13.0 & 0.78$\pm$0.38 & 0.77$\pm$0.35\\
\hline
\multicolumn{5}{c}{\texttt {PCFABS*(POWERLAW+GAUSS})}\\
\hline
&&&\\
\texttt{PCFABS}              &  N$_{\mathrm{H}}$($\times$10$^{22})$ cm$^{-2}$ &$\cdots$   &  $\cdots$  & 2.1$\pm$0.2\\
                             & CvrFract                                      &$\cdots$   &  $\cdots$  & 0.98$\pm$0.01\\
\texttt{POWERLAW}            & $\Gamma$  &  $\cdots$         &$\cdots$     & 2.1$\pm$0.1\\
                             & norm  &  $\cdots$         &$\cdots$& (1.9$\pm$0.1)$\times$10$^{-2}$\\
\texttt{GAUSSIAN}           & LineE  keV     &    $\cdots$   &$\cdots$  & 6.48$\pm$0.20 \\
                            & Sigma  keV     &    $\cdots$   &$\cdots$  & $\leq$0.3\\
                            & norm           &    $\cdots$   &$\cdots$  & (2.0$\pm$1.3)$\times$10$^{-4}$\\
                            & $\chi^{2}/$d.o.f.	  &$\cdots$  &$\cdots$ & 129.25/132\\
&\\
\multicolumn{2}{l}{Unabs. Flux ($\times$10$^{-10}$) \cgs} &$\cdots$ & $\cdots$ & 1.2$\pm$0.1\\
\hline
\end{tabular}\\
\end{table*}

\subsubsection{The \integral\/ data}
\label{INTswift}

Hard X-ray emission from \swift\ is clearly detected in the IBIS 22-60 keV mosaic image, while the source is marginally detected in the JEM-X field of view, at $\sim$5$\sigma$ in the 3.0-10.0 keV energy band. Here we describe the data analysis performed on the \integral\,/IBIS \cite[][]{ubertini03} and \integral\,/JEM-X \cite[][]{lund03} consolidated data of revolution 2144. This data-set has been processed using the standard Off-line Scientific Analysis (OSA v11) software, released by the \integral\/ Scientific Data Centre \cite[][]{courvoisier03}. 

In order to study the X-ray spectral properties of the \swift\ outburst in a broad energy range, we performed a joint spectral fitting by combining together JEM-X and IBIS data, thus covering the 3-200 keV energy band. A systematic error of 2$\%$ was included during the spectral analysis. 

\begin{figure}
   \includegraphics[width=1\columnwidth]{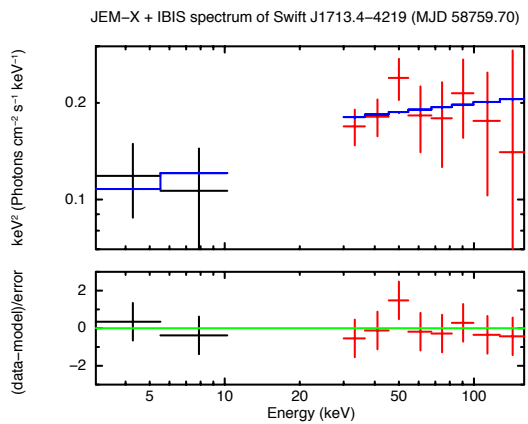}
   \caption{\integral\//JEM-X (in black) and \integral\//IBIS (in red) unfolded spectra (upper panel) together with the residuals in sigma (lower panel). The model used is an absorbed power-law and is plotted as a blue solid line in the upper panel.}
   \label{fig:combspec}
   \end{figure}

In Figure \ref{fig:combspec} we show the resulting JEM-X (black) and IBIS (red) spectrum together with the residuals with respect to the model used for the fit (bottom panel). We model the X-ray emission with an absorbed power-law and we included a constant normalization factor to take into account for the different instrumental calibration. Thus, the resulting model we used is \texttt{CONST*PHABS*POWERLAW}. In Table \ref{tab:Swift1713fit} the best fitting parameters of the JEM-X+IBIS spectral analysis are reported. In particular, we fixed to 1 the constant normalization factor for the IBIS data and we left the JEM-X one free to vary. As the fit appears insensitive to variation in the N$_{\mathrm{H}}$ parameter, we fixed it to 1.04$\times$10$^{22}$ cm$^{2}$, which is the value estimated from the XRT spectral analysis during  the MJD 58779.04 observation (see Sect. \ref{sec:xrtspec} for more details). We obtain a spectral index $\Gamma$=1.90$\pm$0.30 and $\chi^{2}/$d.o.f. of 3.27/7. We obtain only an upper limit for the unabsorbed flux in the 3-200 keV band , which is F$_{3-200} \leq$1.3$\times$10$^{-9}$\cgs.
%@swift1713_abspo_def.xcm in /home/gps/rev_2/WORK/obs/SwiftJ1713_combinedspe

\subsection{The optical/NIR counterpart of \swift}
\label{sec:OPTswift}
%----------------------------------------------------------------------------

The field of \swift\ has been observed by UVOT/{\it Swift} in different epochs and using $U$, $UVW1$ and $UVW2$ filters. In particular, the first two sets of observations have been obtained at MJD 58148.89 and MJD 58149.82, when the source was quiescent, while the last two observations have been performed soon after the end of the new outburst, at MJD 58779.04 and MJD 58785.81, respectively.
In order to investigate the nature of this X-ray transient, we retrieved the available archival {\it Swift} observations and analyzed them. 
To derive the apparent magnitude of the source we used the \texttt{HEAsoft} routine \texttt{uvotsource} with a 5\arcsec\/ wide aperture centered on the position of the star A and a background region of 60\arcsec\/ radius placed in an area free of sources. In Table \ref{tab:UVOTdata} the measured apparent magnitudes in AB system, are reported. 

Although the field was observed by UVOT at different epochs and with different bands, we do not detect the optical counterpart of \swift\ and only moderately deep upper limits are obtained.  

Optical observations of \swift\ have been executed with the LCO telescopes on 2019 October 9, 10 and 12 by \cite{baglio19}. An apparent magnitude of m$_{i}$=(19.738$\pm$0.075) mag has been derived for the star A by applying PSF photometry on the combined $i^{\prime}$ images taken on October 9 and using the APASS catalogue for the calibration. Instead, from the combined images of October 10, the derived apparent magnitude of the star A is m$_{i}$=(20.035$\pm$0.069) mag, showing a variation of 0.3$\pm$0.1 mag with respect to the value derived from the October 9 observations.  

Star A is also found in {\it Gaia} EDR3 and DR2 databases \cite[][]{gaia16, gaia18, riello18, riello20, fabricius20}, with magnitudes $G$ = 20.72, $R_p$ = 19.41 and $B_p$ = 21.64 in EDR3 (reference epoch: 2016.0), and $G$ = 20.80, $R_p$ = 19.31 and $B_p$ = 21.11 in DR2 (reference epoch: 2015.5); we assume that the source was in quescence on both epochs. The attempt to use the {\it Gaia} EDR3 photometric information\footnote{available at https://gea.esac.esa.int/archive/documentation/GEDR3/} to determine the corresponding $i'$ magnitude of \swift\ is however not viable because the relevant {\it Gaia}-to-SDSS transformation holds for 0.5 $< B_p - R_p <$ 2.0, whereas in the present case $B_p - R_p$ = 2.27.

The reported DR2 apparent magnitudes correspond to colours $B_p-R_p$ = 1.80, $B_p-G$ = 0.31 and $G-R_p$ = 1.48 and lie in the range of application of the {\it Gaia}-to-SDSS photometric transformations for DR2; we however note that the signal-to-noise ratio (S/N) for $B_p$ is less than 1.7, which means that formally only a lower limit can be associated to this magnitude. Thus, using the S/N information available in the {\it Gaia} DR2, we determine the actual magnitudes for this object (not corrected for foreground absorption) as reported in Table \ref{tab:Optdata}. 

We therefore applied the photometric transformations from {\it Gaia} passbands to the SDSS photometric system as reported in the {\it Gaia} DR2 documentation\footnote{https://gea.esac.esa.int/archive/documentation/GDR2/} to determine the SDSS $i'$ quiescent magnitude for the source, in order to compare it with the LCO values reported in the literature during outburst. We remark that these conversion formulae depend on the $B_p-R_p$ colour, for which only a lower limit is available. Thus, we evaluated the corresponding $i'$ magnitude by considering a $B_p - R_p$ range determined by the upper limit from the {\it Gaia} data ($>$1.17) on one side and the applicability limit of the transformation formula ($<$4.5) for the case of the SDSS $i'$ photometric band on the other. The average $i'$ magnitude value  obtained in this way, together with the corresponding uncertainty, is reported in Table \ref{tab:Optdata}. For the sake of comparison, we also report there the LCO $i'$-band magnitudes.

\begin{table}
\caption{Results from the UVOT aperture photometry. Different columns correspond to: (1) date of observation; (2) filter; (3) exposure time; (4) apparent magnitudes in AB system.}
\label{tab:UVOTdata}
\centering
\begin{tabular}{llll}
\hline
MJD			&  Filter  & Exp. time & Magnitudes\\
(days)      &               &(s)             & (AB mag)       \\
(1)         & (2)           & (3)            &(4)             \\
\hline		
58148.89    & $U$           &836.6246       & $>$ 20.62 \\
58149.82    & $UVW2$        &930.07         & $>$ 21.25 \\
58779.04    & $UVW1$        &1456.744       & $>$ 21.02 \\
58785.81    & $UVW2$        &1669.948       & $>$ 21.39 \\
\hline
\end{tabular}
\end{table}

\begin{table}
\caption{LCO, {\it Gaia} and VVV DR2 apparent magnitudes for star A}
\label{tab:Optdata}
\centering
\begin{tabular}{lllll}
\hline
Filter	    & LCO	            & LCO             & {\it Gaia} DR2 & VVV DR2\\
            &(2019 Oct 9)       & (2019 Oct 10)   & (2015 June)    & (2010 Mar 29) \\
\hline
 $G$         & $\cdots$         & $\cdots$         & 20.80$\pm$0.02 &  $\cdots$\\
 $B_p$       & $\cdots$         & $\cdots$         & $>$20.48       & $\cdots$\\
 $R_p$       & $\cdots$         & $\cdots$         & 19.31$\pm$0.10 &  $\cdots$\\
& & & &\\
 $g^{\prime}$& $\cdots$         & $\cdots$         & $\cdots$ & $\cdots$\\
 $r^{\prime}$& $\cdots$         & $\cdots$         & $\cdots$ & $\cdots$\\
 $i^{\prime}$& 19.738$\pm$0.075 & 20.035$\pm$0.069 & 20.3$\pm$0.2$^*$ & $\cdots$\\
 & & & & \\
 $J$  & $\cdots$&$\cdots$ &$\cdots$ & 17.843$\pm$0.082 \\
 $H$  & $\cdots$&$\cdots$ &$\cdots$ & 17.165$\pm$0.104 \\
 $K_s$ &$\cdots$ &$\cdots$ &$\cdots$ & 16.738$\pm$0.112 \\
\hline
\multicolumn{4}{l}{$^*$:This magnitude was determined using the {\it Gaia}DR2-SDSS}\\
\multicolumn{4}{l}{photometry transformation formulae (see text).}\\
\hline
\end{tabular}
\end{table}

A NIR counterpart for this source is found in the Second Data Release of the Vista Variables in the V\'ia L\'actea (VVV) survey\footnote{vvvsurvey.org} \cite[][]{minniti10}.
This survey of the Galactic Bulge and inner disk was obtained with the 4.1m VISTA telescope at Cerro Paranal (Chile) between years 2010 and 2015.
The aforementioned source, indicated with the alias VVV J171340.98$-$421838.06, lies 0$\farcs$16 from the optical position of star A reported in \cite{baglio19}, thus well within the positional uncertainty of the latter (0$\farcs$2 according to those authors, while the 1-$\sigma$ positional error of the VVV survey is 0$\farcs$1). The magnitudes in the $JHK_s$ bands are reported in Table \ref{tab:Optdata}.

\begin{figure*}
   \includegraphics[width=2\columnwidth]{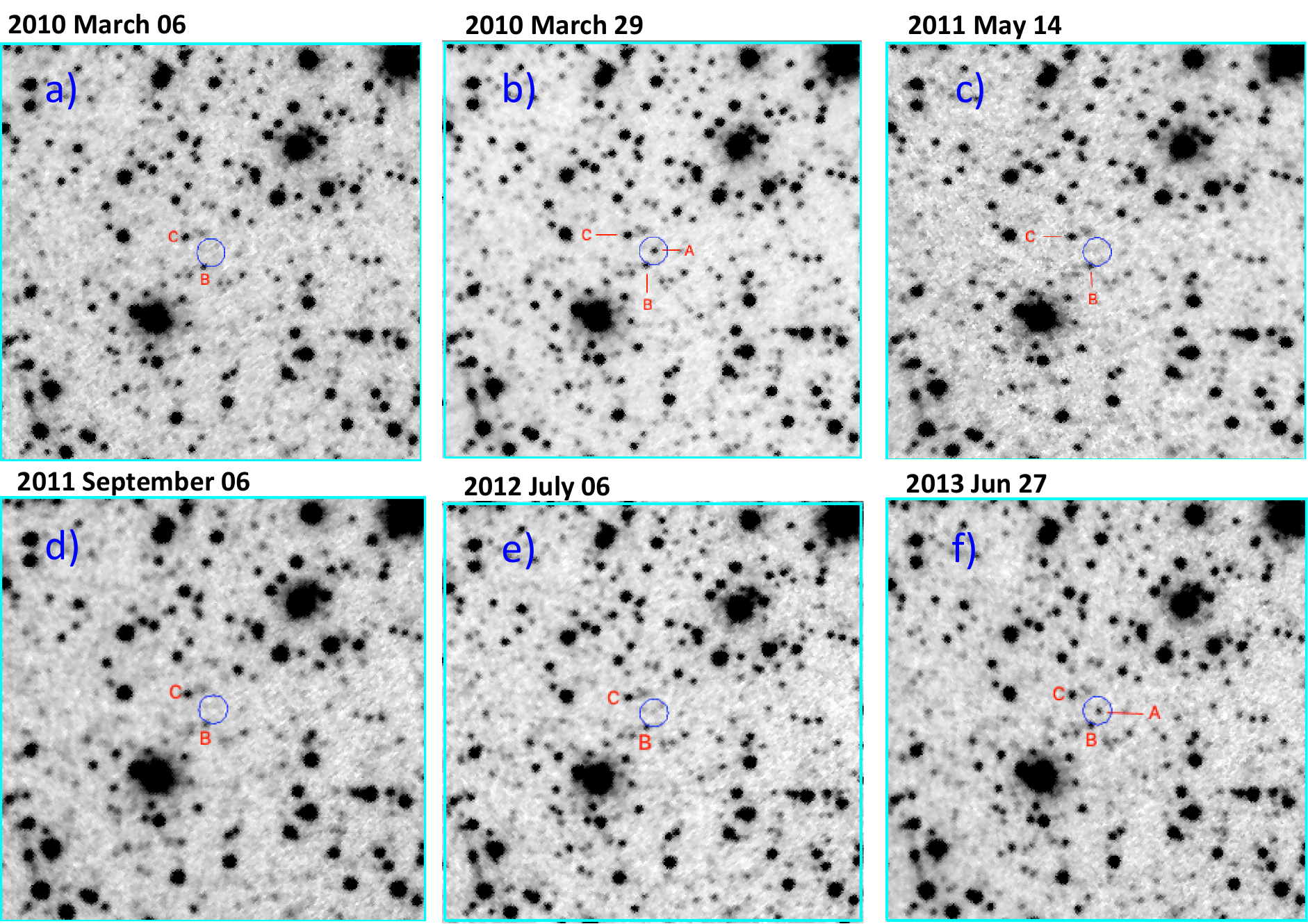}
   \caption{VVV survey $K_s$-band 1$'\times$1$'$ images of the field of view around the XRT error region of \swift\/ (indicated by the circle) acquired at different epochs; North is at top, East to the left. Objects A, B and C are labeled following \citealt{baglio19}. The variability of object A is apparent.}
   \label{fig:VVV_starA}
\end{figure*}

\begin{figure}
   \includegraphics[width=1\columnwidth]{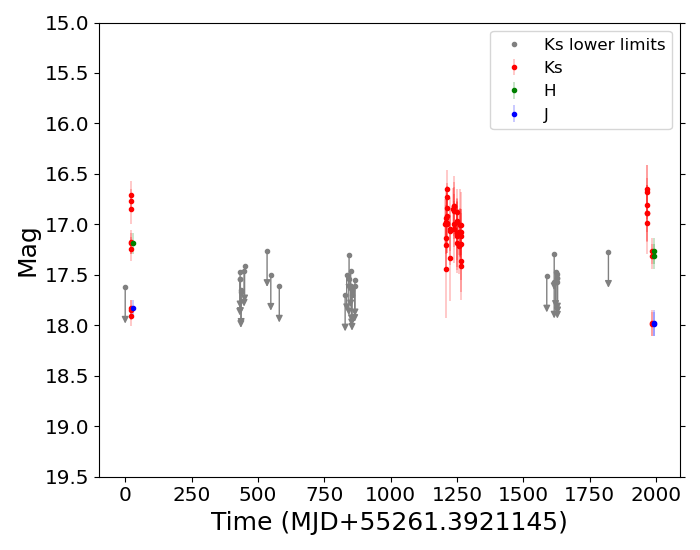}
   \caption{VVV light curve of Swift J1713.4-4219.}
   \label{fig:VVVcurve}
\end{figure}

For the sake of completeness we explored and reduced all the available images containing the field of \swift\/ (tile $d035$ of the survey; details on these NIR observations are reported in the table in the Appendix). We find that the source shows marked variability of at least 1.5 magnitudes in the $K_s$ band over a time scale of months to years (see Figure \ref{fig:VVV_starA} and Figure \ref{fig:VVVcurve}), definitely proving its transient nature and, thus enforcing its identification as the NIR counterpart of \swift.

%\section{Observations and Data analysis}
 %\label{sect_obs}
%-----------------------------------------------------------------------------

\section{Discussion} 
\label{sect_disc}

\subsection{IGR J20155+3827 as a distant supergiant HMXB}

As shown in the previous sections, the multi-wavelength properties of \igr\/, such as the 0.2-10 keV and the optical spectra (see Sections \ref{sec:XMMigr} and \ref{sec:OPTigr}), point to a High Mass X-ray Binary (HMXB) identification.

In particular, although the signal-to-noise is low, the \xmm\/ spectrum in the 0.2-10 keV energy band can be modelled by an absorbed powerlaw with a spectral index $\Gamma$=0.4$^{+0.7}_{-0.9}$, a value which is not unusual in these kinds of source \cite[e.g.][and references therein]{coburn02}.

Moreover, by using the optical dataset available for this source, we may place broad constraints to the spectral type and luminosity class of the secondary star of this system as well as to its distance.

The optical spectrum of the object (Fig. \ref{fig:fig3}) points to an HMXB identification due to the presence of the H$_\alpha$ emission line at a redshift consistent with 0, superimposed on an intrinsically blue spectral continuum. However, the optical spectral shape appears to be largely modified by intervening reddening. This points to the presence of substantial interstellar dust along the source line of sight, which is indeed usual for Galactic HMXBs detected with \integral\/ \cite[see e.g.][]{masetti13} and indicates that the object lies far from the Earth. Moreover, given that the equivalent width of the H$_\alpha$ line (4.5$\pm$0.4 \AA) as measured from the optical spectrum of the source has a value similar to those seen in blue supergiants \cite[][]{leitherer88}, we suggest this broad spectral identification for the companion star in this binary system. Also, the detection of a bright NIR counterpart for IGR J20155+3827 does favour the presence of an early type companion and, consequently, its classification as HMXB.

Unfortunately, the {\it Gaia} EDR3 photometry transformations cannot be used to determine the Johnson magnitudes for this source given that no $B_p$ magnitude is available in the latest data release for it. Similarly, the DR2 information is not applicable as well for this task because the $B_p - R_p$ colour of the object (3.28$\pm$0.01) is out of the range of applicability of the {\it Gaia}-to-Johnson systems photometry transformations, which is $-$0.5 $< (B_p-R_p) <$ 2.75 (see the {\it Gaia} DR2 documentation mentioned in footnote 5 of Section \ref{sec:OPTswift}).

Therefore, we can determine the $V$-band absorption along the line of sight of the \igr\/ by using the USNO-B1.0 catalogue optical magnitudes available for this source \cite[$B \sim$ 17.2--17.5, $R \sim$ 14.1--14.6;][]{monet03} and assuming an intrinsic color of $(B-R)_0 \sim$ $-$0.2 \cite[][]{lang92}, which is typical of early-type stars. In this way we obtain $A_V \approx$ 6.5 mag. This value is comparable, within the above approximations, to the total Galactic absorption towards \igr\/ ($A_V$ = 5.7 mag according to \citealt{schlafly11}) and numerically equal to the estimate of \citealt{schlegel98}. This indeed suggests that the system lies on the far side of the Galaxy with respect to Earth. 

As an aside, we also note that this value is substantially lower than the one derived from the N$_{\rm H}$ parameter obtained in our first X--ray fit (see Table \ref{tab:igr_pnspec}), which is $A_V \approx$ 30 mag (admittedly very large), according to the conversion formula of \cite[][]{predehl95}. This justifies our decision of freezing N$_{\rm H}$ to the Galactic value in our best fit of Section \ref{sec:XMMigr}. 

The above considerations allow us to put general constraints on the distance of the system and, consequently, on the luminosity class of the secondary star. The sky position of \igr\/ corresponds to the Galactic coordinates $l$ = 75$\fdg$90, $b$= +1$\fdg$92. Therefore, taking as a reference the Galaxy map reported in Figure 2 of \cite{bodaghee12} \cite[see also][]{valle08}, it emerges that this object lies in the direction of the Perseus and Cygnus arms of the Galaxy. However, because of the observed reddening, it conceivably lies within or beyond the Cygnus arm, rather than the Perseus one. This implies that the distance to the system is $d \approx$ 8 kpc, which is compatible with the Gaia parallax estimate reported in Section \ref{sec:intro}. 

In turn, given that the de-reddened $R$-band apparent magnitude of the system is $R \sim$ +9.0, its intrinsic absolute optical magnitude is M$_V \approx -5 \div -6$. This further supports the classification of the secondary star as a late O/early B spectral type of luminosity class I, i.e., a blue supergiant. 

\begin{figure}
   \hspace{-1.2cm}
   \includegraphics[width=1\columnwidth,angle=-90]{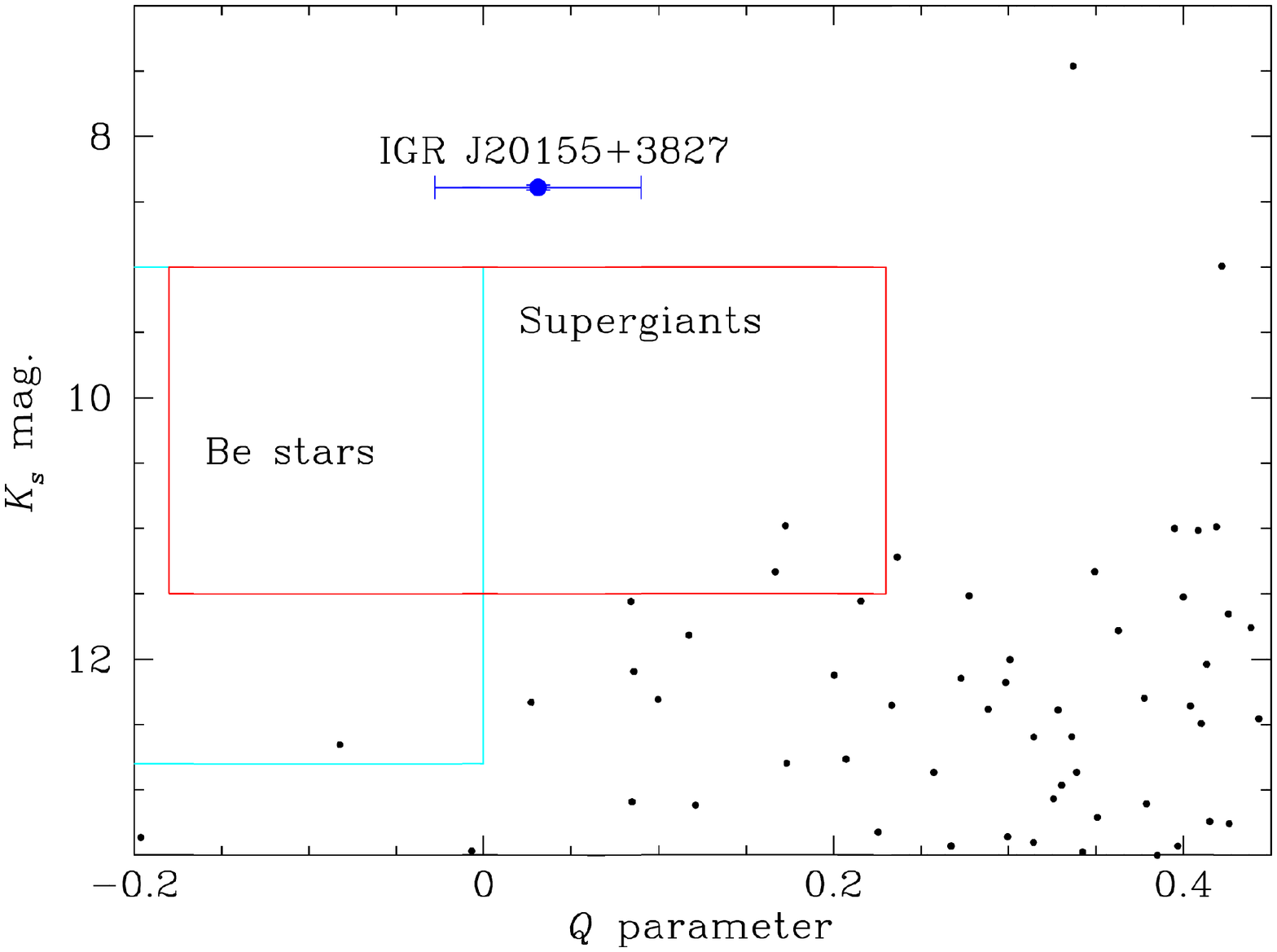}
   \vspace{-1.2cm}
   \caption{$Q$-$K_s$ diagnostic diagram for the field of IGR J20155+3827. Objects falling in a 5$'\times$5$'$ box centered on the position of this high-energy emitter and with accurate 2MASS $JHK_s$ photometry are plotted. The NIR counterpart of the hard X--ray source (blue dot) stands out of the bulk of the field objects and falls close to the region expected to be populated by blue supergiant stars according to \citealt{negueruela07} and \citealt{reig16}. See text for details.}
   \label{fig:qks}
\end{figure}

The spectral type classification proposed here for the optical counterpart of \igr\/ can also be tested through the $K_s$-magnitude vs. $Q$ parameter diagnostic diagram introduced by \cite{comeron05} \cite[see also][]{negueruela07,reig16}. Following these authors, we can use the 2MASS NIR photometry \cite[][]{skrutskie06} to compute the reddening-free parameter $Q$ = $(J-H) - 1.7(H-K_s)$ which, together with the NIR $K_s$ magnitude, allows the construction of a diagram useful to separate early-type from late-type stars. While the latter are mostly concentrated around values $Q$=0.4--0.5 (which correspond to spectral types K to M), the early-type objects typically display $Q \lesssim$ 0. In Figure \ref{fig:qks}, we show the location of \igr\/ in the $Q$-$K_s$ plane, modeled following the right panel of Fig. 1 in \cite{reig16} and constructed by using the NIR photometric information of all 2MASS objects within a box of 5$'\times$5$'$, centered on the position of the optical/NIR counterpart of \igr\/ and for which the magnitudes in the three $JHK_s$ bands are known accurately (i.e. with `AAA' coding in the 2MASS catalogue). From the catalogued 2MASS magnitudes, we obtain a value $Q$ = 0.03$\pm$0.06 for \igr\/, which allows us to place the source in the upper left quadrant of the $Q$-$K_s$ plane of Figure \ref{fig:qks}.
The location of \igr\/ in this diagram falls slightly above the locus of blue supergiant stars in \cite{reig16} but nevertheless in the region populated by this kind of objects, according to \cite{negueruela07}, i.e., with $K_s <$ 11 and $Q <$ 0.2, which supports our conclusions.

We also note that from the unabsorbed flux in the 0.2-10 keV band, as derived from our best fit of the \xmm\/ spectrum of \igr\/, and assuming a distance of d$\sim$7 kpc, we obtain  a luminosity of L${_X}$=(0.3$\pm$0.1)$\times$10$^{34}$ erg s$^{-1}$, whereas that observed with \integral\/ at the outburst peak in the 17-30 keV band roughly corresponds to $\approx$10$^{36}$ erg s$^{-1}$. This indicates an X-ray luminosity dynamic range of $>$100 for IGR J20155+3827.

Finally, this system is not associated with any catalogued radio source and, in the hypothesis of an X-ray binary nature, this implies that it does not display radio-emitting collimated (jet-like) outflows, i.e. this source is not a microquasar.

%\subsection{The LMXB transient nature of Swift J1713.4$-$4219}
\subsection{The X--ray binary nature of Swift J1713.4$-$4219}

The high energy properties of \swift\/ as derived from our analysis of the \integral\/ and {\it Swift} observations, together with the ones inferred from the {\it RXTE}/PCA data reported in \citet[][]{krimm09}, suggest that this source is a X-ray binary. In particular, the emergence of a soft excess and the iron emission line at 6.5 keV in the XRT spectrum of MJD 58785.81 can be indicative of a transition of the source from a hard state, characterized by the emission from the corona, toward a soft state, where the X--ray reflection from an accretion disk produces the iron line at 6.5 keV, which in turn indicates that the line photons are emitted by neutral or low-ionization iron atoms. However, this high energy data-set does not allow us to uniquely derive the nature of the compact object (WD, BH or NS), as the iron line emission is observed both in LMXBs and CVs \cite[e.g.,][]{asai2000,cackett09}.
%The high energy properties of \swift\/ as derived from the \integral\/ and {\it Swift}  observations, suggest that this source is a Low Mass X-ray binary (LMXB) observed to transit from a hard state toward a soft state. In particular, the spectral analysis of the system in the hard state is characterized by the emission from the corona, while in the soft state the X-ray emission is due to an accretion disk with a temperature of T$_{\mathrm{in}}$=1.8$\pm$0.3 keV. The X--ray reflection from the accretion disk produces an iron line at 6.5 keV, which indicates that the line photons are emitted by neutral or low-ionization iron atoms. 

The lower-energy counterpart of \swift\/, star A of \cite{baglio19}, besides being the only optical source inside the subarcsecond {\it Chandra} X--ray error circle, presents substantial variability when one compares the $i'$- and $K_s$-band magnitudes acquired in outburst and in quiescence, reported in Table \ref{tab:Optdata} and in the Appendix.

All this evidence suggests that this object is the actual optical counterpart of \swift. Moreover, the red colours of the source, its low optical/NIR luminosity in quiescence and its location towards the Galactic bulge ($l$ = 345$\fdg$283, $b$ = $-$1$\fdg$986) are compatible with a transient LMXB nature. The large ($\gtrsim$1 mag) excursion of the NIR brightness between outburst and quiescence suggests a (late-type) dwarf star as secondary component of the system.

Assuming a very conservative lower limit $K_s >$ 18 for the quiescent magnitude of \swift, a M0 V spectral type for the companion star in this system, an absolute magnitude of M$_{V} \sim$ +9.0 \cite[][]{lang92}, a $V-K$ = +3.29 color \cite[][]{ducati01} for this classification, and a Galactic absorption $A_{K_s}$ = 0.65 mag \cite[][]{schlafly11} (compatible with the best-fit N$_{\rm H}$ value from the X--ray spectroscopic data of Section \ref{sec:xrtspec}), we obtain a lower limit for distance to the source of $d \gtrsim$ 2.8 kpc. 
Instead, considering an M5 (K0) dwarf as the mass donor star, this limit changes to $d \gtrsim$ 1.4 kpc ($d \gtrsim$ 5.2 kpc). Looking at these numbers, and at the compatibility between our best-fit column density and the total line-of-sight Galactic absorption, we favor an early K dwarf as the secondary star in this binary; this consideration can however be revised once a detection of the quiescent magnitude of the source is achieved.

Finally, from the unabsorbed flux of F$_{3-200}\leq$1.3$\times$10$^{-9}$\cgs, as derived in the spectral analysis of \integral\/ data, and assuming the distance of the system $d \sim$ 5 kpc, we obtain a upper limit for the luminosity in outburst of L$_{3-200 keV} \lesssim$ 4$\times$10$^{36}$erg s$^{-1}$. From the soft state unabsorbed flux in the 0.2-10 keV, we derive a luminosity of L$_{0.2-10 keV}$ = 3.6$\times$10$^{35}$erg s$^{-1}$. These values are fully compatible with what is generally observed in a LMXB system during outburst \cite[see e.g.][]{tanaka96}, but admittedly too large for a CV \cite[e.g.,][]{brunschweiger09,yuasa10,demartino20}.

\section{Conclusion}

We have analysed the multi-wavelength dataset available for two poorly studied X--ray transients, \igr\/ and \swift\/, with the aim to pinpoint their nature. 

From the analysis of the \xmm\/ archival data available for \igr\/ in combination with our Cassini/BFOSC spectroscopic observations of its optical counterpart and the archival optical/NIR photometric data, we have been able to classify this X-ray transient as a distant HMXB. In particular, our analysis implies a distance of the system of $\sim$7 kpc and a blue supergiant classification for the secondary star.  

The data collected for \swift\/, instead, suggest a LMXB transient nature; a CV nature cannot be excluded, albeit this interpretation appears quite unlikely on X-ray luminosity grounds. Indeed, our \integral\/ observations together with the archival XRT dataset have shown the emergence of a soft excess and an iron line at 6.5 keV in the X-ray spectrum taken after the end of the X-ray outburst detected by BAT and IBIS/ISGRI in 2019. Moreover, the observed VVV $K_s$ and LCO $i^{\prime}$ magnitudes obtained for the \swift\/ optical/NIR counterpart, the variability between quiescence and outburst phases, the location towards the Galactic Bulge and the red colours of the source suggest a late-type dwarf classification for the secondary star.

\bigskip

\section*{Acknowledgements}
 We thank the referee for several useful comments that allowed us to improve this paper.
This work has made use of data from the European Space Agency (ESA) mission {\it Gaia} (\url{https://www.cosmos.esa.int/gaia}), processed by the {\it Gaia} Data Processing and Analysis Consortium (DPAC, \url{https://www.cosmos.esa.int/web/gaia/dpac/consortium}). Funding for the DPAC has been provided by national institutions, in particular the institutions participating in the {\it Gaia} Multilateral Agreement.\\
We thank the Loiano Observatory staff (Ivan Bruni, Antonio De Blasi and Roberto Gualandi) for the assistance during the optical observations.\\
We acknowledge the ASI financial/programmatic support via ASI-INAF agreements n.2019-35-HH.0 and n.2017-14-H.0; we also acknowledge the `INAF Mainstream' project on the same subject.\\
F.O. acknowledges the support of the H2020 European Hemera program, grant agreement No 730970, and the support of the GRAWITA/PRIN-MIUR project: "\textit{The new frontier of the Multi-Messenger Astrophysics: follow-up of electromagnetic transient counterparts of gravitational wave sources}". \\ 
This research has made use of the services of the ESO Science Archive Facility and data products from the Vista Science Archive (VSA). Based on observations collected at the European Southern Observatory under ESO programme 179.B-2002 (PI: D. Minniti).

\section*{Data availability}
The data underlying this article are available in the article and in its online supplementary material. The \integral\/ data are publicity available on \url{http://gps.iaps.inaf.it} 

%\newpage
%\begin{figure*}
%   \includegraphics[angle=-90,width=16cm]{1day.eps}
%   \caption{solo per capire dove siamo!}
%   \label{fig2}
%   \end{figure*}

%%%%%%%%%%%%%%%%%%%%%%%%%%%%%%%%%%%%%%%%%%%%%%%%%%



\bibliographystyle{mnras}
\bibliography{mybib_IGR.bib} % if your bibtex file is called example.bib

%\begin{thebibliography}{99}
%\end{thebibliography}

%%%%%%%%%%%%%%%%% APPENDICES %%%%%%%%%%%%%%%%%%%%%

\appendix

\newpage

\section{VVV survey photometry}

\begin{table*}
\caption{Details of VVV survey observations in $JHK_s$ bands of the field of \swift. Different columns correspond to: (1) Date of observations; (2) UT time of observations; (3) MJD of observations; (4) total exposure time; (5) Vega magnitude in $K_s$ band with its 1-$\sigma$ uncertainty, or 3-$\sigma$ magnitude limit for a point source; (6) Vega magnitude in $J$ band; and (7) Vega magnitude in $H$ band.}
\label{tab:VVVswift}
\centering
\begin{tabular}{lllllll}
\hline
Date       & time      & MJD       & exptime     & $K_s$  & $J$  &  $H$  \\ 
           & (UT)      & (days)    & (s)         & (Vega mag) & (Vega mag) & (Vega mag) \\
(1)        & (2)       & (3)       & (4)         & (5)  & (6) & (7)      \\
\hline     
2010-03-06 & 09:22:51  & 55261.39  & 48      &    $>$17.625  & $\cdots$&$\cdots$ \\ % AB to Vega: (18.896 - 1.827) + 0.556
2010-03-29 & 06:39:20  & 55284.28  & 240     &    16.764 $\pm$ 0.112  & 17.830 $\pm$ 0.082 &  17.189 $\pm$ 0.104 \\ % AB(Ks): 19.366; AB(J) to Vega: (19.858 -0.916) + 0.556; AB(H) To Vega: (19.417- 1.366) + 0.556
%%%%% 2011
2011-05-12 & 07:44:33  & 55693.32  & 48     &     $>$17.545  & $\cdots$ & $\cdots$  \\ % AB: 18.816
2011-05-13 & 09:14:49  & 55694.39  & 48     &     $>$17.474  &$\cdots$ &$\cdots$   \\ % AB: 18.745
2011-05-14 & 09:50:33  & 55695.41  & 48     &     $>$17.539  & $\cdots$& $\cdots$  \\ % AB: 18.810
2011-05-16 & 06:56:22  & 55697.29  & 48      &    $>$17.668  &$\cdots$ & $\cdots$  \\ % AB: 18.939
2011-05-17 & 08:22:48  & 55698.35  & 48     &     $>$17.647  &$\cdots$ & $\cdots$  \\ % AB: 18.918
2011-05-28 & 08:03:44  & 55709.34  & 48     &    $>$17.457  & $\cdots$& $\cdots$  \\ % AB: 18.728
2011-05-30 & 07:22:08  & 55711.31  & 44     &     $>$17.414  & $\cdots$&  $\cdots$  \\ % AB: 18.685 
2011-08-23 & 01:03:47  & 55796.04  & 48      &    $>$17.262  & $\cdots$& $\cdots$ \\ % AB: 18.533
2011-09-06 & 02:39:13  & 55810.11  & 48      &    $>$17.497  & $\cdots$&  $\cdots$ \\ % AB: 18.768
2011-10-08 & 00:34:46  & 55842.02  & 48   & $>$17.615  &$\cdots$ &  $\cdots$\\ % AB: 18.886
%%%%% 2012
2012-06-12 & 04:39:57  & 56090.19  & 48  &  $>$17.703  & $\cdots$&$\cdots$\\ % AB: 18.974
2012-06-18 & 01:57:33  & 56096.08  & 48   & $>$17.502  & $\cdots$& $\cdots$ \\ % AB: 18.773
2012-06-25 & 23:47:12  & 56103.99  & 48     &     $>$17.543  & $\cdots$&   $\cdots$   \\ % AB: 18.814
2012-06-27 & 00:41:17  & 56105.03  & 48    &  $>$17.308  &$\cdots$ & $\cdots$\\ % AB: 18.579
2012-07-03 & 01:01:45  & 56111.04  & 48     &     $>$17.464  & $\cdots$&  $\cdots$    \\ % AB: 18.735
2012-07-05 & 23:57:11  & 56114.00  & 48    & $>$17.610  & $\cdots$&$\cdots$  \\ % AB: 18.881
%2012-07-06 & 23:34:28  & 56114.98  & 48   & $>$17.652  & &  \\ % AB: 18.923
2012-07-08 & 01:21:18  & 56116.06  & 48  &  $>$17.695  &$\cdots$ & $\cdots$\\ % AB: 18.966
2012-07-18 & 05:03:31  & 56126.21  & 48   &  $>$17.608  &$\cdots$ &$\cdots$ \\ % AB: 18.879
2012-07-19 & 05:20:43  & 56127.22  & 48   & $>$17.553  &$\cdots$ &$\cdots$  \\ % AB: 18.824
%%%%% 2013
2013-06-24 &  07:06:30  & 56467.30  & 48    & 16.992 $\pm$ 0.244  & $\cdots$&$\cdots$  \\ % AB: 18.841
2013-06-27 & 04:54:27  & 56470.20  & 48  & 17.204 $\pm$ 0.248  & $\cdots$& $\cdots$ \\ % AB: 18.905
2013-06-28 & 02:32:56  & 56471.11  & 48      &  16.942 $\pm$ 0.192  & $\cdots$& $\cdots$\\% AB: 18.929
2013-06-30 & 03:08:19  & 56473.13  & 48      &  16.730 $\pm$ 0.142  & $\cdots$&$\cdots$ \\% AB: 19.056
2013-07-02 & 04:15:05  & 56475.18  & 48  & 16.842 $\pm$ 0.178  &$\cdots$ &  $\cdots$\\ % AB: 18.905
2013-07-11 & 02:21:33  & 56484.10  & 48     & 17.050 $\pm$ 0.211  & $\cdots$& $\cdots$ \\ % AB: 18.956
2013-07-24 & 05:19:26  & 56497.22  & 48      & $>$17.282  & $\cdots$& $\cdots$ \\% AB: 18.553
2013-07-25 & 01:52:29  & 56498.08  & 48      &  16.845 $\pm$ 0.204  & $\cdots$& $\cdots$\\ % AB: 18.823
2013-07-28 & 03:07:49  & 56501.13  & 48      &  16.997 $\pm$ 0.230  & $\cdots$& $\cdots$\\% AB: 18.783
2013-07-28 & 03:56:30  & 56501.16  & 48      &  16.821 $\pm$ 0.189  & $\cdots$&$\cdots$ \\ % AB: 18.819
2013-08-07 & 01:00:32  & 56511.04  & 48      & 17.184 $\pm$ 0.295  & $\cdots$& $\cdots$ \\% AB: 18.700
2013-08-08 & 01:34:34  & 56512.07  & 48      &  16.968 $\pm$ 0.223 &$\cdots$ &$\cdots$ \\% AB: 18.785
2013-08-08 & 04:16:39  & 56512.18  & 48      &  16.981 $\pm$ 0.213  &$\cdots$ &$\cdots$ \\ % AB: 18.860
2013-08-09 & 01:29:53  & 56513.06  & 48      &  16.876 $\pm$ 0.229  &$\cdots$ &$\cdots$ \\ % AB: 18.709
2013-08-09 & 03:19:17  & 56513.14  & 48	  & 17.088 $\pm$ 0.264  & $\cdots$& $\cdots$ \\ % AB: 18.750
2013-08-17 & 00:20:04  & 56521.01  & 48      &  17.216 $\pm$ 0.272  & $\cdots$& $\cdots$\\% AB: 18.832
2013-08-17 & 01:49:43  & 56521.08  & 48      &  17.072 $\pm$ 0.228  & $\cdots$& $\cdots$\\ % AB: 18.887
2013-08-18 & 00:02:16  & 56522.00  & 48     &  17.080 $\pm$ 0.247  & $\cdots$& $\cdots$\\ % AB: 18.805
2013-08-20 & 02:57:00  & 56524.12  & 48      &  17.083 $\pm$ 0.234  & $\cdots$& $\cdots$\\% AB: 18.857
2013-08-21 & 00:34:20  & 56525.02  & 48      &  17.415 $\pm$ 0.336 & $\cdots$& $\cdots$ \\% AB: 18.793
2013-08-21 & 02:36:33  & 56525.11  & 48      & 17.366 $\pm$ 0.301  & $\cdots$&$\cdots$  \\% AB: 18.872
2013-08-22 & 01:53:11  & 56526.08  & 48      &  17.118 $\pm$ 0.232  &$\cdots$ & $\cdots$\\% AB: 18.903
2013-08-23 & 00:58:23  & 56527.04  & 48      &  17.077 $\pm$ 0.228  & $\cdots$& $\cdots$\\% AB: 18.901
2013-09-02 & 00:04:08  & 56537.00  & 48      &  $>$17.413  & $\cdots$&$\cdots$ \\% AB: 18.684
%%%%% 2014
2014-07-12 & 01:17:27  & 56850.05  & 48      &  $>$17.515  & $\cdots$& $\cdots$\\% AB: 18.786
2014-08-08 & 23:42:35  & 56877.99  & 48      &  $>$17.575  &$\cdots$ & $\cdots$\\% AB: 18.846
2014-08-09 & 01:03:40  & 56878.04  & 48      &  $>$17.294  & $\cdots$&$\cdots$ \\% AB: 18.565
2014-08-14 & 01:17:39  & 56883.05  & 48      &  $>$17.469  & $\cdots$&$\cdots$ \\% AB: 18.740
%2014-08-21 & 00:02:00  & 56890.00  & 48      & $>$17.532  & &  \\% AB: 18.803
2014-08-21 & 01:13:44  & 56890.05  & 48      &  $>$17.576  &$\cdots$ &$\cdots$ \\% AB: 18.847
%2014-08-21 & 02:29:09  & 56890.10  & 48      &  $>$17.494  & & \\% AB: 18.765
2014-08-21 & 03:54:22  & 56890.16  & 48      &  $>$17.535  &$\cdots$ &$\cdots$ \\% AB: 18.806	
%%%%% 2015
2015-03-01 & 08:34:36  & 57082.36  & 48      &  $>$17.270  & $\cdots$& $\cdots$\\% AB: 18.541
2015-07-26 & 02:01:12  & 57229.08  & 48      &  16.889 $\pm$ 0.277  &$\cdots$ &$\cdots$ \\% AB: 18.863
2015-07-26 & 02:35:39  & 57229.11  & 48      & 16.682 $\pm$ 0.270  &$\cdots$ & $\cdots$ \\% AB: 18.865
2015-07-26 & 02:51:41  & 57229.12  & 48      &  16.652 $\pm$ 0.236  & $\cdots$& $\cdots$\\% AB: 18.859
2015-07-29 & 03:24:35  & 57232.14  & 48      & $>$17.371  & $\cdots$& $\cdots$ \\% AB: 18.642
%2015- &   & 57248.0354559  & 48      & 17.316456 $\pm$ 0.125257  & &  \\ % dr5
2015-08-14 & 00:51:03  & 57248.04  & 240      & 17.266 $\pm$ 0.130  & 17.985 $\pm$ 0.120 & 17.316 $\pm$ 0.125 \\ % dr5
2015-08-14 & 00:52:10  & 57248.04  & 240      & 17.985 $\pm$ 0.120  & 17.979 $\pm$ 0.129 & 17.265 $\pm$ 0.130 \\ % dr5
%2015- &   & 57248.0412165  & 48      & 17.97889  $\pm$ 0.12869  & &  \\ % dr5
\hline
\end{tabular}\\
\end{table*}



%%%%%%%%%%%%%%%%%%%%%%%%%%%%%%%%%%%%%%%%%%%%%%%%%%

% Don't change these lines
\bsp	% typesetting comment
\label{lastpage}
\end{document}